\documentclass[pra,twocolumn,showpacs,preprintnumbers,amsmath,amssymb,superscriptaddress]{revtex4}
\usepackage{amsmath} 
\usepackage{graphicx} 
\usepackage{braket} 
\usepackage{bm}

\begin{document}

\title{Adiabatic entangling gate of Bose-Einstein condensates based on the minimum function}

\author{Sergi Ortiz}
\affiliation{State Key Laboratory of Precision Spectroscopy, School of Physical and Material Sciences,
East China Normal University, Shanghai 200062, China}
\affiliation{New York University Shanghai, 1555 Century Ave, Pudong, Shanghai 200122, China}
\affiliation{National Institute of Informatics, 2-1-2
Hitotsubashi, Chiyoda-ku, Tokyo 101-8430, Japan}
\affiliation{Polytechnic University of Catalonia, 31 Jordi Girona, 08034 Barcelona, Spain} 

\author{Yilun Song}
\affiliation{New York University Shanghai, 1555 Century Ave, Pudong, Shanghai 200122, China}

\author{June Wu}
\affiliation{Courant Institute of Mathematical Sciences, New York University, New York, NY 10012, USA}
\affiliation{New York University Shanghai, 1555 Century Ave, Pudong, Shanghai 200122, China}

\author{Valentin Ivannikov}
\affiliation{New York University Shanghai, 1555 Century Ave, Pudong, Shanghai 200122, China} 
\affiliation{State Key Laboratory of Precision Spectroscopy, School of Physical and Material Sciences,
East China Normal University, Shanghai 200062, China}

\author{Tim Byrnes}
\email{tim.byrnes@nyu.edu}
\affiliation{New York University Shanghai, 1555 Century Ave, Pudong, Shanghai 200122, China}
\affiliation{State Key Laboratory of Precision Spectroscopy, School of Physical and Material Sciences,
East China Normal University, Shanghai 200062, China}
\affiliation{NYU-ECNU Institute of Physics at NYU Shanghai, 3663 Zhongshan Road North, Shanghai 200062, China}
\affiliation{National Institute of Informatics, 2-1-2 Hitotsubashi, Chiyoda-ku, Tokyo 101-8430, Japan}
\affiliation{Department of Physics, New York University, New York, NY 10003, USA}

\date{\today}

\begin{abstract} 
A scheme is presented to perform an entangling gate between two atomic ensembles or Bose-Einstein condensates in a optical cavity with a common optical mode. The method involves using a generalized Stimulated
Raman Adiabatic Passage (STIRAP) to adiabatically evolve the ground state. We show that dark states exist for any atom number within the cavities, and find that the operation produces an unusual type of evolution where the minimum of the number of atoms between two level transitions to another state.  This produces an unconventional type of entangling Hamiltonian which creates a phase depending on the minimum operation.   We analyze its reliability under a variety of conditions ranging from the ideal decoherence-free case to that including photon loss and spontaneous emission.  Ways of combating decoherence are analyzed and the amount of entanglement that is generated is calculated. 
\end{abstract}


\maketitle
%
%

\section{Introduction} 

Entanglement is one of the defining features of quantum mechanics and is known to be an essential ingredient in performing tasks beyond classical physics, such as  quantum algorithms, quantum metrology, and quantum communication \cite{nielsen00}.  In the context of quantum metrology, multipartite entanglement is used to beat the standard quantum limit where the limits of precision scale as $ \sim 1/\sqrt{N} $, toward the Heisenberg limit which scales as $ \sim 1/N $, where $ N $ is the number of particles in the system \cite{giovannetti11}.  The Heisenberg limit can be approached using entangled states such as NOON states and squeezed states which reduce the noise fluctuations in the signal.  In this context the entanglement properties in an ensemble of qubits has been studied for some time now with many experimental realizations of entangled multipartite states.  For a single ensemble or Bose-Einstein condensate (BEC), entangled states have also been realized where squeezed states and non-Gaussian states have been experimentally realized \cite{riedel10,gerving13,strobel14,pezze16,schmied16}.

\begin{figure}[t]
\includegraphics[width=\linewidth]{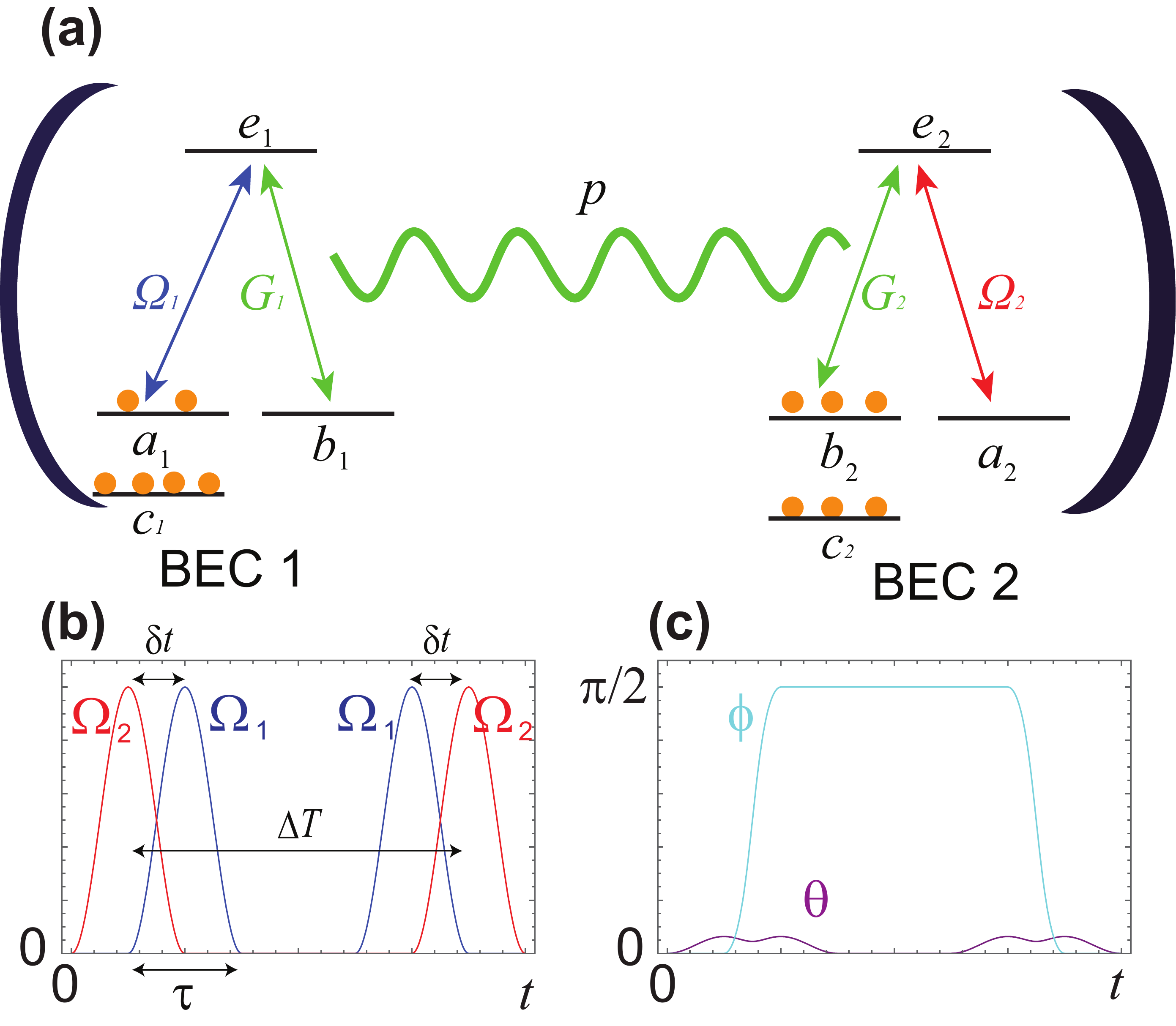} 
\caption{(a) Energy levels and states involved in adiabatic gate for entangling two spinor Bose-Einstein condensates (BECs).  The BECs are placed in an optical cavity, in the strong coupling regime allowing for coherent transfer between the cavities.  The cavity couplings are $G_{1}$, $G_{2}$ between a cavity photon and optical transition between the ground states $b_{i} \leftrightarrow e_{i} $. The classical laser field amplitudes $ \Omega_{i} $ between levels $a_{i} \leftrightarrow e_{i} $ are marked. (b)(c) Typical STIRAP pulse sequence for the two ensemble entanglement.  (b) Amplitudes for the pulses on the two ensembles $ \Omega_1 $ and $ \Omega_2 $. The pulse duration $ \tau $, displacement parameter $ \delta t $ and the time between the two STIRAP sequences $ \Delta T $ are as marked.  (c) The STIRAP pulse sequence in terms of the parametrization $ G_{1,2} = G_0 \cos \theta $, $ \Omega_1 = \Omega_0 \sin \theta \sin \phi $, and $ \Omega_2 = e^{-i\xi} \Omega_0 \sin \theta \cos \phi $. We take $ \Omega_0/G_0 = 0.1 $ here. 
	}
	\label{fig1}
\end{figure}

Multi-ensemble entanglement has been relatively less studied.  Experimentally, entanglement between two ensembles has been pioneered by Polzik and co-workers in the form of a two-mode squeezed state \cite{julsgaard01,sherson08}.  This was used to realize quantum teleportation between two ensembles \cite{krauter13,bao12}. Recently, entanglement between two spatial regions of Bose-Einstein condensates was measured in a single Bose-Einstein condensate \cite{kunkel2017spatially,fadel2017spatial,lange2017entanglement}. To date however, there is no experimental demonstration of entanglement between two independent BECs. Theoretically, there have been many proposals for entanglement generation between ensembles, such as optical cavity mediated methods \cite{pyrkov13,hussain15}, directly interacting the atoms using state-dependent forces \cite{kurkjian13}, and Rydberg excitations \cite{idlas16}.  Many of these proposals generate a $ S^z_1 S^z_2 $ type interaction, where $ S^z_{1,2} $ is the $ z $-component of the total spin for two BECs.  This is known to produce initially a correlation between the $ S^z_1 $ and $S^y_2$  observables between the ensembles, and produce a ``Devil's crevasse'' entanglement structure \cite{byrnes13,kurkjian13}. Such multi-ensemble entanglement could be used for the purposes of quantum metrology \cite{Gross2012} and quantum information processing \cite{byrnes15,byrnes12}.

In this paper, we introduce a method of entangling two BECs or ensembles, which gives rise to an unconventional effective interaction. 
This is performed using a geometric phase technique, where a stimulated Raman adiabatic passage (STIRAP) is applied onto two ensembles mediated by photons. The geometric or Berry phase \cite{berry84} is acquired in addition to the dynamic phase in an adiabatic system that 
undergoes a closed path and has been extensively studied for general purposes and adiabatic quantum computation \cite{moller08}. The STIRAP method consists of an adiabatic and robust transfer of particles between ground states without
populating the excited states \cite{bergmann98,shore90}. This is advantageous in terms of spontaneous emission as this is one of the serious decoherence channels for Bose-Einstein condensates in particular, which is enhanced by a factor of  $N $ due to superradiance.  

This paper is organized as follows: In section \ref{sec:stirap} we describe the general procedure to generate entanglement in our scheme. As the type of entanglement that is generated has not been analyzed before in the literature, we describe the nature of the entanglement.  In section \ref{sec:mingate} we show the features of the phase gate without decoherence, using a STIRAP and geometric gate scheme.  
In section \ref{sec:numerics} we test the robustness of the scheme by direct numerical simulation. Finally, in section \ref{sec:conc} we summarize our findings.

\section{Optical cavity entanglement using STIRAP}
\label{sec:stirap}

\subsection{Experimental configuration}

In this section we describe the our scheme for generating entanglement between two ensembles. 
We consider that the ensembles are placed in a optical cavity and coupled with optic fibers in a similar configuration to that 
discussed in Ref. \cite{pyrkov13}.  Each of the atoms are assumed to possess several ground states that can be used to store quantum information, labeled by $ a_i, b_i, c_i $, where the labels on the ensemble are $ i = 1,2 $.  The pairs $ (a_1,c_1) $ and $ (b_2,c_2) $ are the states that are used as the ``logical'' states, and have relatively long storage and decoherence times.  The remaining levels $ b_1,a_2,e_i $ are used for the purposes of the entangling operation, and do not necessarily have to have good storage or decoherence properties as they are only populated for short durations of time.  For example, for $^{87}\text{Rb} $ the logical states may be the magnetically trapped hyperfine states $ a_1 = | F = 1, m_F = -1 \rangle $ and $ c_1 = | F = 2, m_F = 1 \rangle $ [34-36] and the second ground state are $ b_1 =| F = 2, m_F = 2 \rangle $ for the first ensemble.  For the second ensemble, we may use $ a_2 = | F = 1, m_F = -1 \rangle $, $ b_2 = | F = 2, m_F = 1 \rangle $, and $ c_2= | F = 2, m_F = 0 \rangle $.  
An excited state $ e_i $ is available on each ensemble which allows for a Raman transition between levels $ a_i $ and $ b_i $.  The cavity is coupled to the transition $ b_i \leftrightarrow e_i $, and produces or absorbs a common cavity photon mode $ p $.  Strong coupling of optical photons to ensembles and BECs have
been realized in a variety of configurations, varying from cavities in magneto optical traps (MOTs) to atom chip systems \cite{colombe07,walther06,kimble08,duan00,abdelrahman14}.  The transition $ a_i \leftrightarrow e_i $ is controllable via a polarized laser, and is assumed to be controllable in terms of the pulses that can be generated.  

The basic idea of our scheme is then to perform a STIRAP sequence with the Raman transition between levels $ a_i $ and $ b_i $, in the presence of the coupled cavity system. While the configuration has similarities to previous works such as Ref. \cite{pyrkov13}, we will see that this produces an unconventional type of effective interaction Hamiltonian between the two ensembles with rather different properties. The Hamiltonian of the above scheme is
\begin{align}
H  = &H_{\text{las}} + H_{\text{cav}} + H_{\Delta}\label{eq:ham} \\
H_{\text{las}}= & \hbar \sum_{i=1,2} \Omega_{i}(t)(a_{i}^{\dag} e_{i}+e_{i}^{\dag} a_{i})  \\
H_{\text{cav}}= & \hbar\sum_{i=1,2}  G_{i}(b_{i}^{\dag}e_{i} p^{\dag}+p e_{i}^{\dag}b_{i})  \\
H_{\Delta}= & \hbar \Delta_{e}(e_{1}^{\dag}e_{1}+e_{2}^{\dag}e_{2}) 
\end{align}
where $ p $ is a common photonic mode that is shared between the ensembles.  A common photonic mode can be approximated if 
the coupling between the cavity photons and the photon mode in the fiber is sufficiently strong as shown in Ref. \cite{pyrkov13}.  Alternatively, the two ensembles may be placed within the same cavity. In this case, the BECs should be positioned in spatial points where the cavity field is comparable.   The parameter $ \Omega_{1,2} $ is the amplitude of the laser fields inducing a transition between $ a_i \leftrightarrow e_i $, $G_{1,2}$ is
the atom cavity mode coupling, and $ \Delta_e $ is the detuning between the excited
and the ground states for both the cavity and the laser pulses.

A typical pulse sequence for the laser fields is shown in Fig. \ref{fig1}(b). We assume a functional form of the pulses to take a form 
\begin{align}
\Omega_1(t) & = A(t-\delta t ) + A(t-\Delta T+\delta t) \nonumber \\
\Omega_2(t) & = e^{-i\omega t} \left( A(t) + A(t-\Delta T) \right)
\label{pulsedef}
\end{align}
where a single pulse of length $ \tau $ and amplitude $ \Omega_0 $ takes a form
\begin{align}
A(t) \equiv \left\{
\begin{array}{cc}
\Omega_0 \sin^2 (\frac{\pi t}{\tau}) & 0 \le t \le \tau \\
0 & \text{otherwise} 
\end{array}
\right.
\end{align}
and $ \omega $ is the frequency variation of the $ \Omega_2 $ pulse. As is typical of STIRAP sequences, a ``counter-intuitive'' sequence is used, where first the laser with zero population is switched on.  The second laser pulse is then turned on corresponding to where the atoms are populated.  We point out that unlike a standard STIRAP pulse, the lasers are applied on {\it different} ensembles.  Due to the presence of the cavity, the excitations (defined as either an atom in $ e_i $ or a photon) couple to quantum states that link these two states.

\subsection{Dark states: qubit case}

We now show that dark states are present in the Hamiltonian as given in (\ref{eq:ham}), which will justify the STIRAP entangling procedure.  We first derive explicit expressions for the dark states for qubits  $ N_1 = N_2 = 1 $, which will help to introduce the more general ensemble case in the next section.  The scheme in this case reduces to that introduced in Ref. \cite{amniat05}. We assume that initially the state is prepared on the long-lived logical states $ c_1, a_1, c_2, b_2 $. For concreteness, let us say the initial state is
\begin{align}
| \psi(t=0) \rangle & = \frac{1}{2} \left( | c_1 \rangle + | a_1 \rangle \right) \left( | c_2 \rangle + | b_2 \rangle \right)  \nonumber \\
& = \frac{1}{2} \left( | c_1  c_2 \rangle +  | c_1 b_2 \rangle +  | a_1  c_2 \rangle +   | a_1 b_2 \rangle  \right) \label{initstate}
\end{align}
which is an unentangled state.  The aim of the procedure will be to generate entanglement between the qubits. Each of the terms in (\ref{initstate}) follow a different time evolution under the Hamiltonian (\ref{eq:ham}).  We now discuss the effect on each of the terms.  

For the term $  | c_1  c_2 \rangle $, the Hamiltonian performs no operation on this state as it is completely decoupled from both the laser transitions and the cavity coupling.  Likewise, the state $ | c_1 b_2 \rangle $ undergoes no evolution as in order to excite the state $ b_2 $ to $ e_2 $ a cavity photon is required, and none are present.  Writing the projection operators $ P_{c_1 c_2} = |c_1 c_2 \rangle \langle c_1 c_2|$ and $ P_{c_1 b_2} = |c_1 b_2 \rangle \langle c_1 b_2|$, the projection of the Hamiltonian are 
\begin{align}
H_{c_1 c_2} & = P_{c_1 c_2} H P_{c_1 c_2} =0 \\
H_{c_1 b_2} & = P_{c_1 b_2} H P_{c_1 b_2} =0
\end{align}
and hence there is no time evolution of this particular state. 

For the term $ | a_1  c_2 \rangle$, the laser on atom 1 can create an excitation to state $ e_1 $, which can in turn transition to $ b_1 $ with the emission of a cavity photon.  On atom 2, there is no effect as the state of the atom is in the decoupled state $ c_2 $.  Thus we may write the Hamiltonian in the space of the states $\{ |a_1 0 \rangle, |e_1  0  \rangle, |b_1 1 \rangle \} $ as
\begin{align}
H_{a_1 c_2} = P_{a_1 c_2} H P_{a_1 c_2}= \hbar \left(
\begin{array}{ccc}
0 & \Omega_1 & 0  \\
\Omega_1  & \Delta_e & G_1 \\
0 & G_1  & 0 
\end{array} \right)  ,
\end{align}
where second index labels the photon Fock states
\begin{align}
| l \rangle = \frac{1}{\sqrt{l!}} \left( p^\dagger \right)^l | 0 \rangle  
\end{align}
and $ P_{a_1 c_2} $ is defined as 
\begin{align}
P_{a_1 c_2}& = \ |a_1 0 \rangle \langle a_1 0|+|e_1 0 \rangle \langle e_1 0| + |b_1 1 \rangle \langle b_1 1|.
\end{align}
This Hamiltonian has an eigenstate with zero energy, i.e. a dark state, of the form
\begin{align}
|D_1 \rangle = \frac{1}{\sqrt{{\cal N}_1}} \left( G_1 |a_1  0 \rangle - \Omega_1  |b_1  1  \rangle\right)  .
\label{d1dark}
\end{align}
where $ {\cal N}_1 $ is a suitable normalization constant. Such a state which does not involve any excited states can be used in a STIRAP procedure.  Initially when the laser is off, $ \Omega_1 = 0 $ and the ground state is simply $ |a_1 0 \rangle $.  When the laser is turned on, the state adiabatically follows (\ref{d1dark}), until the laser is turned off again, where it returns to $ |a_1 0 \rangle $.  

For the term $ | a_1 b_2 \rangle $, the Hamiltonian in this case may be written as
\begin{align}
H_{a_1 b_2} = P_{a_1 b_2}H P_{a_1 b_2}= \hbar \left(
\begin{array}{ccccc}
0 & \Omega_1 & 0 & 0 & 0 \\
\Omega_1 & \Delta_e & G_1 & 0 & 0  \\
0 & G_1 & 0 & G_2 & 0  \\
0 & 0 & G_2 & \Delta_e & \Omega_2  \\
0 & 0 & 0 & \Omega_2 & 0 
\end{array} \right) ,
\end{align}
where $ P_{a_1 b_2} $ is defined as 

\begin{align}
P_{a_1 b_2}& = \ |a_1 b_2 0 \rangle \langle a_1 b_2 0|+|e_1 b_2 0 \rangle \langle e_1 b2_0| + |b_1 b_2 1 \rangle \langle b_1 b_2 1| \nonumber  \\
& +\ |b_1 e_2 0 \rangle \langle b_1 e_2 0|+|b_1 a_2 0 \rangle \langle b_1 a_2 0|.
\end{align}

This has a dark state
\begin{align}
|D_2 \rangle = \frac{1}{\sqrt{{\cal N}_2}} \left( G_1 \Omega_2  |a_1 b_2  0\rangle - \Omega_1 \Omega_2 |b_1 b_2 1\rangle + G_2 \Omega_1 
|b_1 a_2 0 \rangle \right)
\label{qubitdarkd2}
\end{align}
where $ {\cal N}_2 $ is a suitable normalization constant. At the beginning of the STIRAP evolution, only $ \Omega_2  $ is turned on and hence the dark state is $ |D_2 \rangle  =  |a_1 b_2  0\rangle $.  The STIRAP procedure then adiabatically evolves this state such that finally $ \Omega_1  $ is turned on, and $ \Omega_2 = 0 $, which corresponds to $ |D_2 \rangle  =  |b_1 a_2 0 \rangle $.  Thus in this case there is a transition which swaps the ground states $ a_i \rightarrow b_i $ and $ b_i \rightarrow a_i $.  

Now that we have derived the effect of the first STIRAP pair, let us consider the effect of the second STIRAP pair.  Clearly as this is the same operation but time-reversed, this will simply evolve the states back to their original configuration.  
We may thus summarize the effect of the STIRAP sequence (dropping the photon number $ l $ which are zero everywhere)
\begin{align}
 & \text{STIRAP 1}&  & \text{STIRAP 2} & \nonumber \\
 | c_1  c_2 \rangle & \hspace{5mm} \longrightarrow &  | c_1  c_2 \rangle & \hspace{5mm} \longrightarrow & | c_1  c_2 \rangle  \nonumber \\
 | c_1 b_2 \rangle & \hspace{5mm} \longrightarrow & | c_1 b_2 \rangle &  \hspace{5mm} \longrightarrow & | c_1 b_2 \rangle \nonumber \\
 | a_1  c_2 \rangle & \hspace{5mm} \longrightarrow  & | a_1  c_2 \rangle & \hspace{5mm} \longrightarrow &   e^{i\gamma_1} | a_1  c_2 \rangle\nonumber \\
 | a_1 b_2 \rangle  & \hspace{5mm} \longrightarrow  & | b_1 a_2 \rangle & \hspace{5mm} \longrightarrow  &  e^{i\gamma_2} | a_1 b_2 \rangle ,
\label{phaserelation}
\end{align}
where we have added phases $ \gamma_{1,2}$ to the evolutions of the $ | a_1  c_2 \rangle , | a_1  b_2 \rangle  $ states as there is a Berry phase due to the adiabatic evolution.  These will be derived in Sec. \ref{sec:effham}.

\begin{figure*}[t]
		\includegraphics[width=\linewidth]{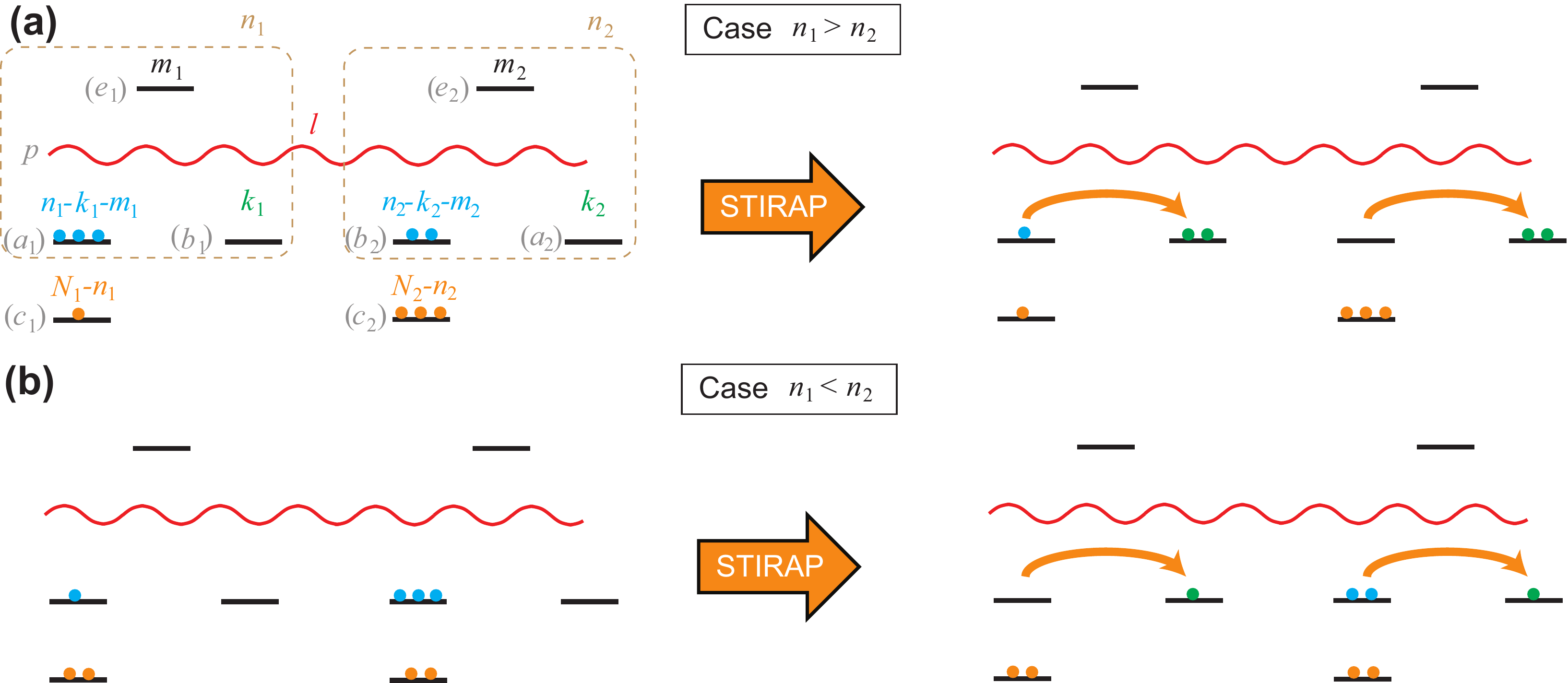} 
	\caption{Effect of STIRAP pulses for Fock states. The labeling convention for Fock states as given in (\ref{fockdef}) is given.  Labels in brackets refer to the associated bosonic operators, while remaining labels give the number occupying each state in the Fock basis. (a) The effect of the STIRAP pulses for $ n_1 > n_2 $.  This shows the particular case of $ N_1 = 4, N_2=5, n_1=3, n_2=2$.  The initial state has $ k_1=k_2= 0 $ which transitions to $ k_1=k_2= \text{min} (n_1,n_2) = 2 $.  (b) The effect of the STIRAP pulses for $ n_1 < n_2 $.  This shows the particular case of $ N_1 = 3, N_2=5, n_1=1, n_2=3$.  The initial state has $ k_1=k_2= 0 $ which transitions to $ k_1=k_2= \text{min} (n_1,n_2) = 1 $. 	In both cases the excited states $m_{1,2} $ are unoccupied during the STIRAP sequence. The cavity photon number $ l $ increases during the procedure but is zero at the beginning and the end. }
	\label{fig1b}
\end{figure*}

\subsection{Dark states: ensemble case}
\label{sec:ensemblecase}

We now show an explicit form of the dark states for the ensemble case.   We assume firstly that the number of atoms in each ensemble is fixed, and no particle loss occurs throughout the process.  This implies that 
\begin{align}
N_i = c_i^\dagger c_i + a_i^\dagger a_i  + b_i^\dagger b_i + e_i^\dagger e_i
\end{align}
where $ N_i $ for $ i=1,2 $ are constants. Observing from Fig. \ref{fig1}(a) that the levels $ c_i $ are completely decoupled from the transitions. Hence throughout the STIRAP operation the numbers on the $c_i $ and remaining levels is conserved as follows:
\begin{align}
& n_i  = a_i^\dagger a_i  + b_i^\dagger b_i + e_i^\dagger e_i \nonumber \\
& N_i - n_i  = c_i^\dagger c_i  .
\end{align}
We may thus consider each sub-particle number space $ n_i $ separately.  This is the same procedure as the previous section where we considered the four terms in (\ref{initstate}) separately. 

Let us now establish some notation for the basis states to describe the quantum state of the system. Define the Fock states of the system as
\begin{align}
& |k_1, m_1, k_2, m_2, l \rangle_{n_1 n_2} = \nonumber \\
& \frac{(a_1^\dagger)^{n_1-k_1-m_1} (b_1^\dagger)^{k_1}  (a_2^\dagger)^{k_2} (b_2^\dagger)^{n_2-k_2-m_2}}{\sqrt{(n_1-k_1-m_1)! k_1! k_2! (n_2-k_2-m_2)! }}
\nonumber  \\
& \times \frac{ (e_1^\dagger)^{m_1} (e_2^\dagger)^{m_2} (p^\dagger)^l (c_1^\dagger)^{N_1-n_1} (c_2^\dagger)^{N_2-n_2}}{\sqrt{m_1! m_2! l! (N_1-n_1)! (N_2-n_2)!}}| 0\rangle ,
\label{fockdef}
\end{align}
where the number of particles in each state are labeled according to Fig. \ref{fig1b}.  We label the $ n_i $ variables as subscripts as they are effectively conserved numbers throughout the STIRAP evolution and hence the dynamics only involve the change of the remaining labels.  The initial state of the system before the STIRAP sequence is described as some superposition of states where only the states $ a_1, c_1, b_2, c_2 $ are occupied, with zero photons in the cavity, implying $ k_i = m_i = l = 0 $.  It then follows that the initial state wavefunction is written
\begin{align}
|\psi (0) \rangle = \sum_{n_1=0}^{N_1} \sum_{n_2=0}^{N_2} \psi_{n_1 n_2} |0, 0, 0, 0, 0 \rangle_{n_1  n_2} .
\label{geninitstate}
\end{align}
From such an initial state, during the STIRAP operation the photon number $ l $ becomes determined entirely by the $ k_i $ parameters.  In the first ensemble, one cavity photon is emitted for each atom that is present in level $ b_1 $, from $ H_{\text{cav}} $ in (\ref{eq:ham}).  Similarly, every atom that leaves level $ b_2 $ reduces the cavity photon by one.  This means that given a starting state with zero cavity photons, the number is fixed to 
\begin{align}
l = k_1 - k_2 - m_2.
\end{align}

We may now construct the dark state in terms of these basis states.  A dark state by definition only involves the ground states of the atoms, which are the levels $ a_i, b_i, c_i $.  Let us again find the dark state for a particular sub-particle number sector $n_1,  n_2 $ (i.e. each term in  (\ref{geninitstate})).  Assuming that initially there are no photons, the dark state should involve the states
\begin{align}
|D \rangle_{n_1 n_2} = \sum_{k_1=0}^{n_1} \sum_{k_2=0}^{k_1}  D_{k_1 k_2}^{(n_1 n_2)} |k_1, 0, k_2, 0, k_1-k_2 \rangle_{n_1 n_2}
\label{generaldark}
\end{align}
where $D_{k_1 k_2}^{(n_1 n_2)} $ are coefficients to be determined. Here we note that we require $ k_1 \ge k_2 $ as the photon number cannot be negative.  As a dark state should be an eigenstate of a Hamiltonian with zero eigenvalue, applying (\ref{eq:ham}) to the above should result in all terms canceling exactly.  This allows us to obtain the wavefunction analytically. Let us first apply the Hamiltonian to a Fock state involved in the dark state (\ref{generaldark}):  
\begin{align}
H & |k_1, 0, k_2, 0, k_1-k_2 \rangle_{n_1 n_2}  = \nonumber \\
&  G_1 \sqrt{(k_1-k_2)k_1} |k_1-1, 1, k_2, 0, k_1-k_2-1 \rangle_{n_1 n_2} \nonumber \\
+ & G_2 \sqrt{(k_1-k_2)(n_2-k_2)} |k_1, 0, k_2, 1, k_1-k_2-1 \rangle_{n_1 n_2} \nonumber \\
 + & \Omega_1 \sqrt{n_1-k_1} | k_1, 1, k_2, 0, k_1-k_2 \rangle_{n_1 n_2} \nonumber \\
 + & \Omega_2 \sqrt{k_2} | k_1, 0, k_2-1, 1, k_1-k_2 \rangle_{n_1 n_2}.
\end{align}
Using this we may find the effect of applying $ H $ to (\ref{generaldark}) directly, which gives
\begin{align}
H  |D \rangle_{n_1 n_2} = & \nonumber \\
\sum_{k_1=0}^{n_1} \sum_{k_2=0}^{k_1} &  \Big[ \big( D_{k_1+1 k_2}^{(n_1 n_2)}  G_1  \sqrt{(k_1-k_2+1)(k_1+1)} \nonumber \\
+ &  D_{k_1 k_2}^{(n_1 n_2)} \Omega_1 \sqrt{n_1-k_1} \big) |k_1, 1, k_2, 0, k_1-k_2\rangle_{n_1 n_2}  \nonumber \\
+ & \big( D_{k_1 k_2-1}^{(n_1 n_2)}  G_2 \sqrt{(k_1-k_2+1)(n_2-k_2+1)} \nonumber \\
+ & D_{k_1 k_2}^{(n_1 n_2)} \Omega_2 \sqrt{k_2} \big) |k_1, 0, k_2-1, 1, k_1-k_2 \rangle_{n_1 n_2}  \Big],
\end{align}
where we have shifted the indices by one for the terms proportional to $ G_i $.  Setting the coefficients for the states to zero gives us the conditions required for a dark state:
\begin{align}
\frac{D_{k_1+1 k_2}^{(n_1 n_2)}}{D_{k_1 k_2}^{(n_1 n_2)}} & = -\frac{\Omega_1 \sqrt{n_1 - k_1}}{G_1 \sqrt{(k_1-k_2+1)(k_1+1)}} \nonumber \\
\frac{D_{k_1 k_2+1}^{(n_1 n_2)}}{D_{k_1 k_2}^{(n_1 n_2)}} & = - \frac{G_2 \sqrt{(k_1-k_2)(n_2-k_2)}}{\Omega_2 \sqrt{k_2+1}} .
\end{align}
Using these relations it is possible to find a closed expression for the coefficients of the  dark states, which is given by 
\begin{align}
D_{k_1 k_2}^{(n_1 n_2)} = \frac{(-1)^{k_1 + k_2}}{\sqrt{\mathcal N}}  \left( \frac{\Omega_1}{G_1} \right)^{k_1}  \left( \frac{G_2}{\Omega_2} \right)^{k_2}
\sqrt{ \frac{{n_1 \choose k_1} {n_2 \choose k_2}}{(k_1-k_2)!}} ,
\label{darkstateformula}
\end{align}
where $ {\mathcal N} $ is a suitable normalization constant and only terms with $ k_1 \ge k_2  $ and $ 0\le k_i \le n_i $ are defined.  

Let us now examine the limiting behavior of this dark state.  Initially $ \Omega_2 $ is turned on, and $ \Omega_1 = 0 $. Due to the $ \Omega_1^{k_1} $ term, the only non-zero terms are those with $ k_1 = 0 $.  Furthermore, since we require $ k_1 \ge k_2 $, this also sets $ k_2 = 0 $. Thus the dark state initially for $  \Omega_1/\Omega_2 \rightarrow 0 $ is
\begin{align}
|D \rangle_{n_1 n_2} = |0, 0, 0, 0, 0 \rangle_{n_1 n_2} \hspace{1cm}  \Omega_1/\Omega_2 \rightarrow 0.
\end{align}
Next consider the reverse limit where $ \Omega_2 \ll \Omega_1 $.  Furthermore, let us work in a regime such that $ \Omega_1,\Omega_2  \ll G_1, G_2 $. First let us examine the case when $ n_1 > n_2 $.  In this case the largest coefficient in (\ref{darkstateformula}) is obtained by making the $  \left( \frac{G_2}{\Omega_2} \right)^{k_2} $ term large, which is achieved by making the $ k_2 $ coefficient as large as possible which corresponds to $ k_2 = n_2 $.  Since $ k_1 \ge k_2 $, the valid range of $ k_1 $ is $ n_2 \le k_1 \le n_1 $.  Due to the $  \left( \frac{\Omega_1}{G_1} \right)^{k_1} $ term, the largest coefficient favors small $ k_1 $, hence in this limit the state approaches $ k_1= k_2 = n_2 $.  For $ n_2 > n_1 $, due to the $  \left( \frac{G_2}{\Omega_2} \right)^{k_2} $ term again the largest coefficient is obtained by making $ k_2 $ large.  However, since $ k_1\ge k_2 $ and $ k_1 $ can only be $ n_1 $ at most, the largest term is $ k_2 = n_1 $.  To maximize the $  \left( \frac{\Omega_1}{G_1} \right)^{k_1} $ term, again small $k_1 $ is needed, so in this case the state approaches $ k_1 = k_2 = n_1 $.  In summary, after the the STIRAP evolution the minimum of the total sub-particle number evolves to levels $ b_1 $ and $a_2 $ (see Fig. \ref{fig1b}):
\begin{align}
|D &  \rangle_{n_1 n_2} = \nonumber \\
& |\text{min}(n_1,n_2), 0, \text{min}(n_1,n_2), 0, 0 \rangle_{n_1 n_2} \hspace{1cm} \Omega_2/\Omega_1 \rightarrow 0 .
\end{align}
Here we have not kept track of the phases which evolve on each state.  This will be examined in the next section.

\subsection{Berry phases due to STIRAP evolution}
\label{sec:effham}

We have shown in the previous sections that the action of the STIRAP is to move the minimum of the number of atoms that populates levels $ a_1 $ and $ b_2 $ to levels $b_1$ and $ a_2 $ respectively. This can be the basis of an entangling gate based on geometric phases that are produced by the adiabatic process.  In this section we derive the phases that are produced by this process, and derive an effective entangling Hamiltonian.

Let us first examine the case of qubits, to understand the basic mechanism of the phases that are produced.  Consider again the  four terms in (\ref{initstate}) which evolve separately throughout the STIRAP pulse.  For the states $ |c_1 c_2 \rangle $ and $ | c_1 b_2 \rangle $, the Hamiltonian (\ref{eq:ham}) has no effect on the states hence there is no phase that is picked up during the evolution.  For the state $|a_1 c_2 \rangle $, we have established that the dark state is
\begin{align}
|D_1 \rangle = \cos \theta |a_1   0\rangle  + \sin \theta | b_1  1 \rangle 
\end{align}
where we have parameterized $ G_1 = \cos \theta $ and $ \Omega_1 = \sin \theta $ and assumed that there is no phase difference between them.  In this case the Berry phase is 
\begin{align}
\gamma_1  = i \int_{\theta_i}^{\theta_f} d \theta \langle D_1 | \frac{\partial}{\partial \theta} | D_1 \rangle
= 0
\label{gammaone}
\end{align}
since the integrand is zero. 

Turning now to the remaining term $|a_1 b_2 \rangle $, we may write the dark state in this case as
\begin{align}
|D_2 \rangle = & \frac{2}{\sqrt{\sin^2 2 \theta + \sin^4 \theta \sin^2 2 \phi}} \Big[  \cos \theta \sin \theta \cos \phi | a_1 b_2 \rangle \nonumber \\
& - \sin^2 \theta \sin \phi \cos \phi| b_1 b_2 \rangle 
 + e^{i\xi} \cos \theta \sin \theta \sin \phi |b_1 a_2 \rangle  \Big]
\end{align}
where we have parameterized 
\begin{align}
G_{1,2} & = G_0 \cos \theta \nonumber \\
\Omega_1 & = \Omega_0 \sin \theta \sin \phi \nonumber \\
\Omega_2 & = e^{-i\xi} \Omega_0 \sin \theta \cos \phi .
\label{parameterize}
\end{align}
We work in the regime where $ G_0 \gg \Omega_0 $, hence with these parameters during the STIRAP pulse $ \theta \ll 1 $. During the first STIRAP pulse initially $ \Omega_2/\Omega_1 =0 $, which then changes to $ \Omega_1/\Omega_2 = 0 $, and finally reverts back to $ \Omega_2/\Omega_1 =0 $, hence $ \phi $ changes from $ 0 \rightarrow  \pi/2 \rightarrow 0 $ (see Fig. \ref{fig1b}(d)).  The Berry phase for this state is
\begin{align}
\gamma_2  = & i \int_{\theta_i}^{\theta_f} d \theta \langle D_2 | \frac{\partial}{\partial \theta} | D_2 \rangle
+ i \int_{\phi_i}^{\phi_f} d \phi \langle D_2 | \frac{\partial}{\partial \phi} | D_2 \rangle \nonumber \\
& + i \int_{\xi_i}^{\xi_f} d \xi \langle D_2 | \frac{\partial}{\partial \xi} | D_2 \rangle
\label{berrydef2}
\end{align}
Evaluating the integrand for the first two terms in (\ref{berrydef2}) give zero and the third term gives
\begin{align}
\gamma_2  = -\int_{\xi_i}^{\xi_f} d \xi 
\frac{\sin^2 \phi \sin^2 2 \theta }{\sin^2 2 \theta + \sin^4 \theta \sin^2 2 \phi} .
\end{align}

As $ \theta,\phi,\xi $ are all time dependent, they are all implicitly dependent on each other.  From (\ref{pulsedef}) the phase on $ \Omega_2 $ takes a form 
\begin{align}
\xi = \omega t
\end{align}
We obtain 
\begin{align}
\gamma_2  \approx -\omega \int_{t_i}^{t_f} dt \sin^2 \phi(t) 
\label{gammatwoint}
\end{align}
where we have used the fact that $ \theta \ll 1 $.  Let us further assume that the time between the pulses in (\ref{pulsedef}) are much longer than that of the pulse durations $ \Delta T \gg \tau $.  In this case, the dominant part of the integral in (\ref{gammatwoint}) is during the two STIRAP pulses and we may approximate
\begin{align}
\gamma_2  \approx - \omega (\Delta T + \tau)
\label{finalphase}
\end{align}
where we have taken the time duration between the STIRAP to be the point where $ \phi(t) = \pi/4 $.
Eq. (\ref{finalphase}) has the simple interpretation that the state picks up a phase $ \gamma_2  $ according to how long level $ a_2 $ is occupied.  This gives rise to an entangling gate for the qubit case as shown in (\ref{phaserelation}) as the only phase that is picked up is on the state $a_1 b_2 $.  

For the general ensemble case, we obtain similar results. Using the same parameterization as  (\ref{parameterize}), substituting into (\ref{darkstateformula}) gives
\begin{align}
D_{k_1 k_2}^{(n_1 n_2)} = & \frac{(-1)^{k_1 + k_2}}{\sqrt{\mathcal N}} 
\sqrt{ \frac{{n_1 \choose k_1} {n_2 \choose k_2}}{(k_1-k_2)!}} \nonumber \\
& \times e^{i \xi k_2} \tan^{k_1-k_2} \theta  
\sin^{k_1} \phi \cos^{-k_2} \phi .
\end{align}
In the regime of $ G_0 \gg \Omega_0 $, we have $ \theta \ll 1 $ and we can say that the largest terms come from $ k_1 = k_2 $
\begin{align}
| D \rangle \approx \frac{1}{\sqrt{\mathcal N}} \sum_{k=0}^{\min(n_1,n_2)} e^{i \xi k} \tan^k \phi 
\sqrt{{n_1 \choose k_1} {n_2 \choose k_2}}
\end{align}
where the upper limit on the sum comes from similar arguments as that at the end of Sec. \ref{sec:ensemblecase}.  Evaluating the Berry phase (\ref{berrydef2}) using this expression for the dark state we obtain
\begin{align}
\gamma(n_1,n_2) = - \frac{\omega}{\mathcal N} \int_{t_i}^{t_f} dt \sum_{k=0}^{\min(n_1,n_2)} k \tan^{2k} \phi {n_1 \choose k_1} {n_2 \choose k_2}
\end{align}
where the normalization in this case is $ {\mathcal N} = \sum_{k=0}^{\min(n_1,n_2)} \tan^{2k} \phi {n_1 \choose k_1} {n_2 \choose k_2} $. Interpreting the $ \tan^{2k} \phi {n_1 \choose k_1} {n_2 \choose k_2}/{\mathcal N}  $ as a probability distribution, we see that this is strongly peaked at $ k=0 $ for $ \phi = 0 $ and the maximal value $ k = \min(n_1,n_2) $ when $ \phi = \pi/2 $.  This therefore gives the same basic effect as for the qubit case, where the Berry phase is picked up between the two STIRAP pulses.  We may therefore approximate
\begin{align}
\gamma(n_1,n_2)= -\omega \min(n_1,n_2) (\Delta T+\tau ).
\end{align}
We thus see that a phase is picked up on states depending upon the minimum of sub-particle sector in which the state is.  For the special case of qubits, $ n_1, n_2 = 0,1 $ hence we see that the only case that a phase is picked up is when $ n_1 = n_2 = 1 $.  This corresponds to the state $ | a_1 b_2 \rangle $, and agrees with (\ref{gammaone}) and (\ref{finalphase}).  We summarize the general phase transformation of the two STIRAP pulses according to
\begin{align}
& |0,0,0,0,0\rangle_{n_1,n_2} \rightarrow | \min(n_1,n_2),0, \min(n_1,n_2),0,0 \rangle_{n_1,n_2}  \nonumber \\
& \rightarrow  e^{i\gamma(n_1,n_2)} |0,0,0,0,0\rangle_{n_1,n_2}
\end{align}
Taking the whole operation together, one may write an effective Hamiltonian for the process
\begin{equation}
\label{eq:exoticH}
H_{\text{eff}}/\hbar  = \omega \min(n_{1},n_{2})
\end{equation}
which is evolved for a time $ \Delta T + \tau $.

\section{Entanglement properties of the miniumum gate}
\label{sec:mingate}

We now discuss some of the basic properties of the entangled state that is produced by the effective Hamiltonian derived in the previous section. The Hamiltonian (\ref{eq:exoticH}) produced by the adiabatic procedure produces a phase depending on the minimum of the number of bosons occupying levels $ a_1 $ and $ b_2 $.  Let us consider initially preparing the state in a $ S^x $ eigenstate with respect to the logical states
\begin{align}
|\psi(t=0)\rangle  = | \frac{1}{\sqrt{2}}, \frac{1}{\sqrt{2}} \rangle \rangle_1 | \frac{1}{\sqrt{2}}, \frac{1}{\sqrt{2}} \rangle \rangle_2 
\label{initcondition}
\end{align}
where the spin coherent states are defined as 
\begin{align}
| \alpha, \beta \rangle \rangle_1 & = \frac{1}{\sqrt{N_1!}} \left( \alpha a^\dagger_1 + \beta c^\dagger_1 \right)^{N_1} | 0 \rangle \label{spincoherentstate1} \\
& = \frac{1}{\sqrt{N_1!}}  \sum_{n_1 = 0}^{N_1} \sqrt{{N_1 \choose n_1} } \alpha^{n_1} \beta^{N_1 - n_1} | n_1 \rangle
\label{spincoherentstate2} 
\end{align}
for the first ensemble and 
\begin{align}
| \alpha, \beta \rangle \rangle_2 & = \frac{1}{\sqrt{N_2!}} \left( \alpha b^\dagger_2 + \beta c^\dagger_2 \right)^{N_2} | 0 \rangle \label{spincoherentstate3} \\
& = \frac{1}{\sqrt{N_2!}}  \sum_{n_2 = 0}^{N_2} \sqrt{{N_2 \choose n_2} } \alpha^{n_2} \beta^{N_2 - n_2} | n_2 \rangle
\label{spincoherentstate4} 
\end{align}
for the second ensemble. Here we defined the Fock states
\begin{align}
| n_1 \rangle & = \frac{1}{\sqrt{n_1 ! (N_1 - n_1)!}} (a_1^\dagger)^{n_1} (c_1^\dagger)^{N_1-n_1}  |0 \rangle \nonumber \\
| n_2 \rangle & = \frac{1}{\sqrt{n_2 ! (N_2 - n_2)!}} (b_1^\dagger)^{n_2} (c_2^\dagger)^{N_2-n_2}  |0 \rangle .
\end{align}

According to the discussion of the previous section, after the two STIRAP pulses, the state evolves to
\begin{align}
\label{eq:wFTime}
|\psi(t)\rangle = &\frac{1}{\sqrt{2^{N_1+ N_2}}}
\Big( \sum_{n_{2}=0}^{N_2} \sum_{n_{1}= n_{2}}^{N_1}   \sqrt{\binom{N_1}{n_{1}}\binom{N_2 }{n_{2}}}  e^{i n_{2} t}  |n_{1}n_{2}  \rangle \\
& + \sum_{n_{1}=0}^{N_1} \sum_{n_2 = n_1+1}^{N_2}   \sqrt{\binom{N_1}{n_{1}}\binom{N_2 }{n_{1}}}  e^{i n_{1}t}|  n_{1}n_{2}  \rangle  \Big) ,
\end{align}
where the phase depending on $ \min (n_1,n_2) $ was used. 

\begin{figure}
		\includegraphics[width=\linewidth]{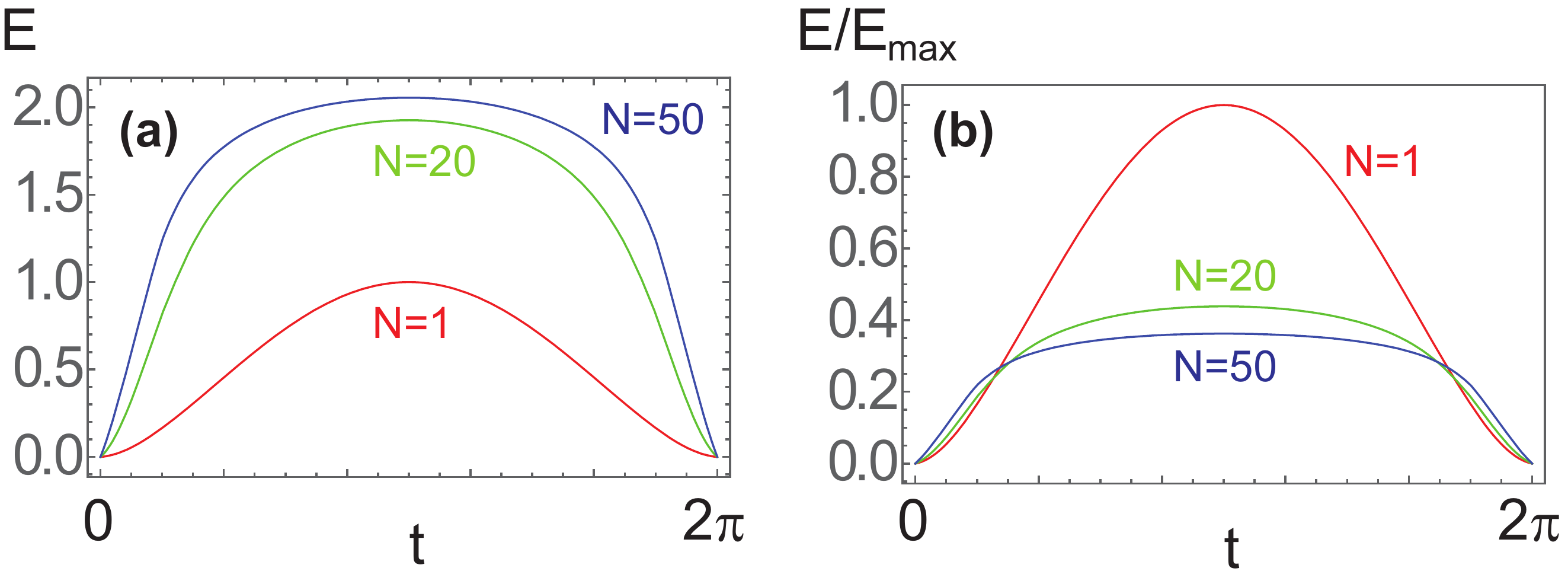} 
	\caption{Entanglement produced by the minimum gate as a function of entangling time. Subfigures show (a) the total amount of entanglement and (b) the normalized entanglement relative to the maximally entangled state $ E_{\text{max}} = \log_2 (N+1) $. We assume particle numbers as marked and $ N_1 = N_2 = N $. }
	\label{fig2}
\end{figure}

First let us verify that an entangled state is produced by the gate.  For a pure bipartite ensemble system as we consider here, the von Neumann entropy 
\begin{align}
E = - \rho_1 \log_2 \rho_1 = - \sum_j \lambda_j \log_2 \lambda_j
\end{align}
quantifies the entanglement, where $ \rho_1 $ is the density matrix with a partial trace taken over ensemble 2
\begin{align}
\rho_1 = \text{Tr}_2 |\psi(t)\rangle \langle \psi(t) | = \sum_{n_2=0}^{N_2} \langle n_2 | \left( 
|\psi(t)\rangle \langle \psi(t) |  \right) | n_2 \rangle 
\end{align}
and $ \lambda_j $  are the eigenvalues of  $ \rho_1 $.  In Fig. \ref{fig2}(a) we show the entanglement generated by the minimum gate.  We see that entanglement is generated between the ensembles, with the maximum value occurring at $ t = \pi $, with a periodicity of $ t = 2\pi $.  The amount of entanglement increases with particle number, which is expected as the dimensionality of the systems increase with particle number, allowing for a larger capacity of entanglement.  For qubits $ N_1 = N_2 = 1 $, a Bell state is produced at $ t = \pi $, which is a maximally entangled state.  For larger ensembles, the gate does not produce a maximally entangled state (Fig. \ref{fig2}(b)).  One example of a maximally entangled state is
\begin{align}
| \psi_{\text{max}} \rangle = \frac{1}{\sqrt{N+1}} \sum_{n=0}^{N} | n  n \rangle  
\label{maxentangle}
\end{align}
which has an entanglement equal to 
\begin{align}
E_{\text{max}} = \log_2 (N+1)
\label{maxent}
\end{align}
where we have assumed that $ N_1 = N_2 = N $. Other maximally entangled states can be produced by local operations on (\ref{maxentangle}). The minimum gate cannot produce such maximally entangled states, but still produce significant amounts of entanglement between the ensembles.  The type of entanglement is a non-local variety, as opposed to entanglement between particles within the same ensemble, as has been observed to date in BECs  \cite{Gross2012,schmied16}.   However, unlike $ S^z_1 S^z_2 $ interactions which produce a complex ``devil's crevasse'' structure in the entanglement \cite{byrnes13}, this interaction produces a smooth increase and decrease in the entanglement.

We may also analyze the type of state that is produced via the $Q$-functions, which plots a quasi-probability distribution according to the overlap with spin coherent states.  Due to the two ensembles involved, in general the $Q$-function involves four real variables $ (\theta_1, \phi_2, \theta_2, \phi_2) $ corresponding to the parametrization
\begin{align}
\alpha_1 & = \cos \tfrac{\theta_1}{2} \nonumber \\
\beta_1 & = \sin \tfrac{\theta_1}{2} e^{i \phi_1} \nonumber \\
\alpha_2 & = \cos \tfrac{\theta_2}{2} \nonumber \\
\beta_2 & = \sin \tfrac{\theta_2}{2} e^{i \phi_2}
\label{alphabeta}
\end{align}
and overlaps are taken with the spin coherent states (\ref{spincoherentstate1}).   For the sake of visualization, we therefore plot the states where the projection is taken on various $ S^z_2 $ eigenstates:
\begin{align}
P_{n_2} = | n_2 \rangle \langle n_2 | . 
\end{align}
We thus define the conditional $Q$-function as
\begin{align}
Q_{n_2} (\theta_1, \phi_1) =  \langle \langle \alpha_1, \beta_1 | 
P_{n_2} |\psi(t)\rangle \langle \psi(t) | P_{n_2}  | \alpha_1, \beta_1  \rangle \rangle ,
\label{condqfunc}
\end{align}
where the parameterization (\ref{alphabeta}) is implicit.  We may also define the marginal $Q$-function where the partial trace is taken over one of the ensembles
\begin{align}
Q_1 (\theta_1, \phi_1) = \langle \langle \alpha_1, \beta_1 | \rho_1 | \alpha_1, \beta_1  \rangle \rangle .
\label{marginalqfunc}
\end{align}

\begin{figure}
		\includegraphics[width=\linewidth]{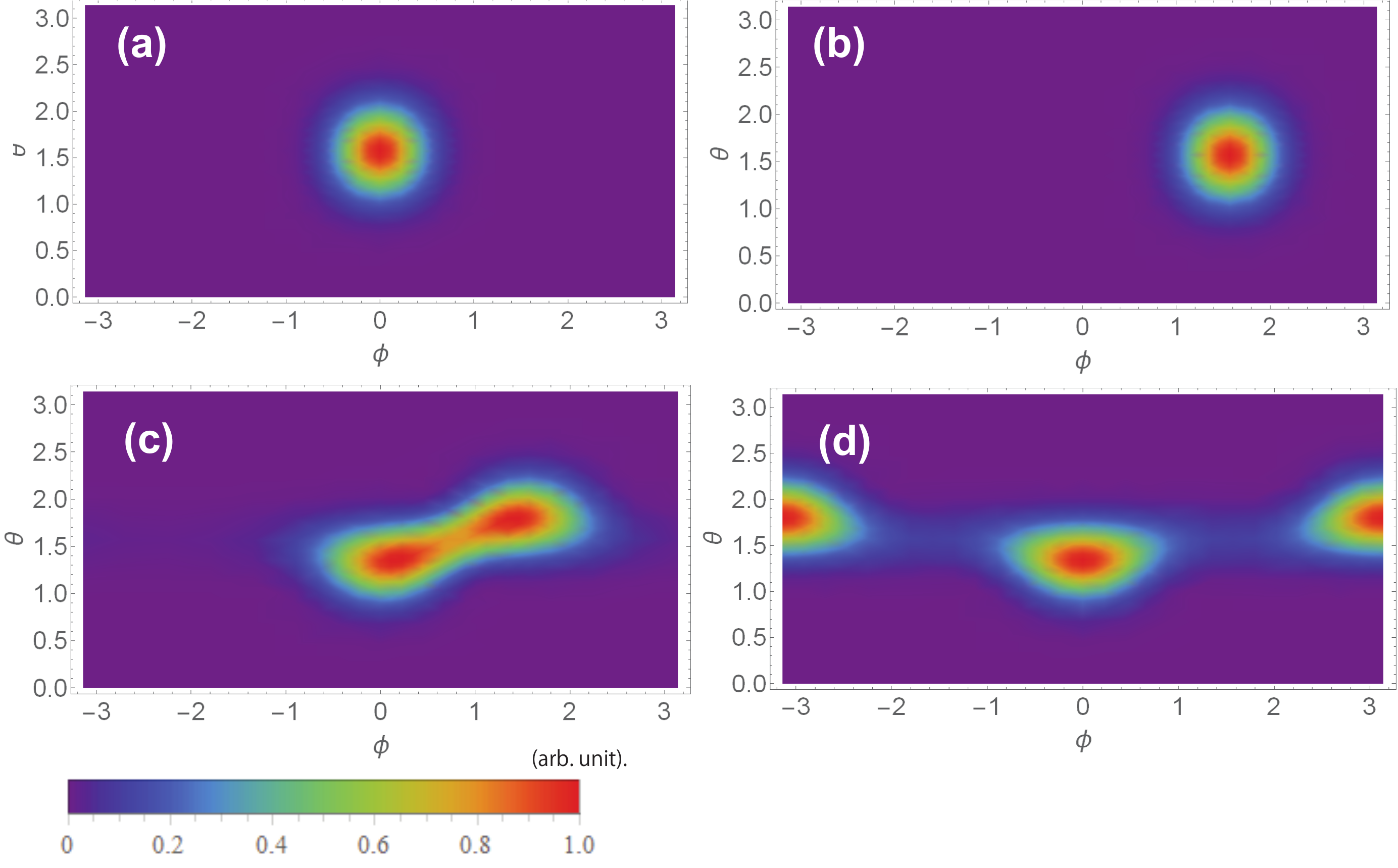} 
	\caption{Conditional $Q$-functions as defined in (\ref{condqfunc}).  The minimum gate is evolved for parameters (a) $ t = \pi/2 $, $n_2 = 0 $; (b)  $ t = \pi/2 $, $n_2 = N $; (c) $ t = \pi/2 $, $n_2 = N/2 $;  (d) $ t = \pi $, $n_2 = N/2 $.  Particle numbers $ N_1 = N_2 = 20 $ are used for all plots. }
	\label{fig3}
\end{figure}

Figure \ref{fig3} shows the conditional $Q$-function for various choices of $ n_2 $, the projection parameter. We can interpret each of the graphs as being the particular type of state that a given $ | n_2 \rangle $ state is entangled with.  For a choice $ n_2 = 0 $, the state on ensemble 1 is unchanged from the initial state. It is a $ S^x_1 $ eigenstate centered around $ \theta_1 = \pi/2 , \phi_1 = 0 $.  On the other hand, for $ n_2 = N_2 $, the state is rotated around the equator of the Bloch sphere by an angle equal to $ t $, the interaction time.  At intermediate $ n_2 $, there is a combination of the two effects, where the Gaussian is ``sliced'' into two parts, determined by the $ n_2 $ chosen.  The upper half of the Gaussian rotates by an angle $ t $, whereas the lower half is left unrotated.  

\begin{figure}
		\includegraphics[width=\linewidth]{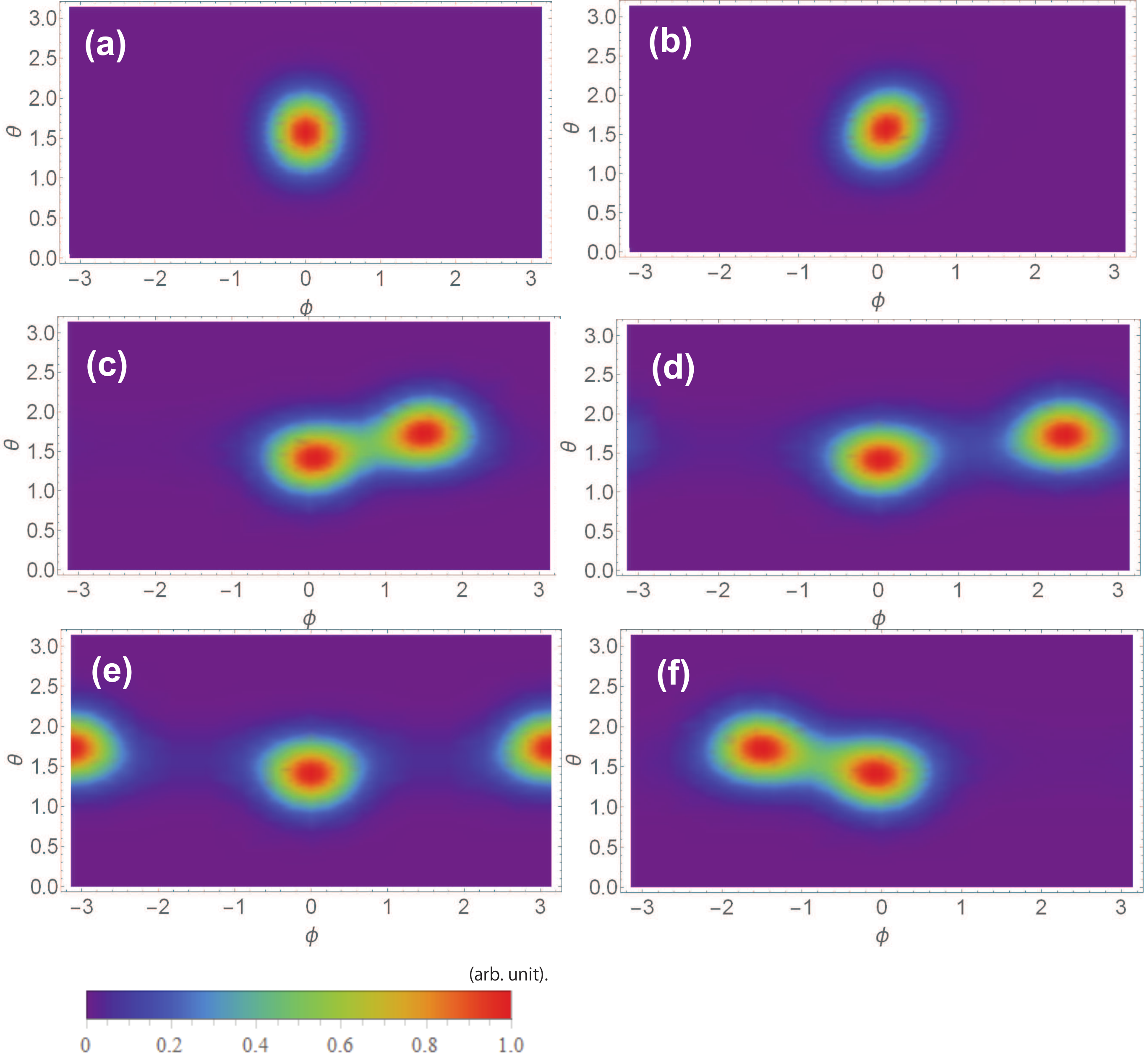} 
	\caption{Marginal $Q$-functions as defined in (\ref{marginalqfunc}).  The minimum gate is evolved for parameters (a) $ t = 0 $; (b)  $ t = 1/\sqrt{N} $; (c) $ t = \pi/2 $; (d) $ t = 3\pi/4 $;  (e) $ t = \pi $;  (f) $ t = 3\pi/2 $. 	Particle numbers $ N_1 = N_2 = 20 $ are used for all plots. }
	\label{fig4}
\end{figure}

The marginal $Q$-functions give distributions which give a probabilistic sum of the conditional $Q$-functions, weighted by the probabilities (Fig. \ref{fig4}).  The typical $Q$-distribution appears located in two locations.  The upper half of the Gaussian rotates on average an angle $ t $, while the lower half remains on average in the same position. The distributions have a non-Gaussian form for entangling times that rotate the distributions to a significant extent, which occur for times $ t \gtrsim 1/\sqrt{N} $.  We may thus say that the minimum gate produces entangled states with highly non-Gaussian
characteristics.

\section{Numerical evolution of STIRAP gates}
\label{sec:numerics}

In order to demonstrate the theory of the previous sections, we numerically time evolve the Hamiltonians for small system sizes.  In a real experimental situation decoherence will be inevitably present.  In our scheme the most important decoherence channels to consider are spontaneous emission of the atoms from the excited state $ e_i $ to the ground states, and photon loss.  However, taking this into account greatly increases the numerical overhead of the simulation, as it becomes necessary to simulate the evolution of a density matrix, instead of a wavefunction.  For this reason we simulate both the case with and without decoherence, first to analyze the adiabadicity of the STIRAP gates, then to see the robustness of the gates under decoherence.

For the case not involving decoherence, we may simply evolve the Schrodinger equation as the state is always pure. The coherent evolution of our state under STIRAP is described by
\begin{equation}
\label{schrodinger}
\frac{d \psi_{\bm{n} }}{dt} = -\frac{i}{\hbar} \sum_{\bm{n}'} H_{  \bm{n} \bm{n}'} \psi_{\bm{n}'} 
\end{equation}
where we define 
\begin{align}
 | \bm{n} \rangle & = | k_1, m_1, k_2, m_2, l \rangle_{n_1 n_2}  \nonumber \\
 \psi_{\bm{n} } & = \langle \bm{n}| \psi \rangle \nonumber \\
H_{  \bm{n} \bm{n}'}  & = \langle \bm{n}| H | \bm{n}' \rangle
\label{statedef}
\end{align}
To numerically solve (\ref{schrodinger}) we diagonalize the matrix $ H_{  \bm{n} \bm{n}'} $ and obtain the state at time $ t $ according to
\begin{align}
\psi_{\bm{n} } = \sum_{\bm{n}'} e^{-i \epsilon_{\bm{n}'} t/\hbar} \langle \bm{n} |  \epsilon_{\bm{n}'} \rangle \langle  \epsilon_{\bm{n}'} | \psi(0) \rangle 
\end{align}
where $ |  \epsilon_{\bm{n}} \rangle $ is the eigenstate with eigenvalue $ \epsilon_{\bm{n}} $ and the initial state $ | \psi(0) \rangle $ is given by  (\ref{geninitstate}).  

To include the effects of spontaneous emission and photon loss, we must evolve the master equation with the corresponding Lindblad terms, written as
\begin{equation}
\label{eq:Lindblad}
\frac{d\rho}{dt}=-\frac{i}{\hbar}[H,\rho]+\frac{\Gamma_{s}}{2}\sum_{i=1,2}(\mathcal{D}[F_{a,i}^{-}]\rho+\mathcal{D}[F_{b,i}^{-}]\rho)+\frac{\Gamma_{\gamma}}{2}\mathcal{D}[c]\rho ,
\end{equation}
where the basis states of the density matrix $\rho$ are taken to be Fock states (\ref{fockdef}) and the Hamiltonian $H$ is given by equation (\ref{eq:ham}). In the master equation, the first term describes the coherent evolution of the system, and the second term describes spontaneous emission where we assume that the excited state $|e_{i}\rangle$ decays to both $|a_{i}\rangle$,  $|b_{i}\rangle$ with the same rate for simplicity. The last term describes photon loss through the mirrors at a rate $\Gamma_{\gamma}$. The Lindblad superoperator is
\begin{align}
\mathcal{D}[O]\rho & \equiv 2 O \rho O^{\dag}-O^{\dag}O \rho-\rho O O^{\dag}
\end{align}
for an arbitrary operator $ O $ and we have defined
\begin{align}
F_{a,i}^{-} & \equiv a_{i}^{\dag}e_{i} \nonumber \\
F_{b,i}^{-} & \equiv b_{i}^{\dag}e_{i}.
\end{align}
To solve (\ref{eq:Lindblad}) we use a numerical differential equation solver in Mathematica with the initial state 
\begin{align}
\rho(t=0) = |\psi(t=0) \rangle \langle \psi(t=0) |
\end{align}
where the state is given in (\ref{initcondition}).  We note that it has recently been found that the dephasing for the ac Stark shift using a non-Markovian calculation has been found to be greatly suppressed \cite{lone15}.  Thus although we make a Markovian assumption here, in practice the amount of decoherence could be less than what is estimated in our simulations.  

Due to the number of levels involved, the dimension of the Hilbert space quickly increases with boson numbers $ N_{1,2} $.  To make the problem more tractable, we tried to use several approximations to reduce the dimensionality. As our approach is to use an adiabatic transition to evolve the state along the ground state, we would like to effectively capture the dark states of the system.  From (\ref{generaldark}), the dark state should only involve Fock states taking the form $ | k_1, 0, k_2, 0, k_1- k_2 \rangle $, which requires $ k_1 \ge k_2 $ and has zero population of the excited and photon states.   We may thus take these and their adjacent states to reduce the computational overhead. For example, the excited states can be limited to states $ l,m_{1,2} \in [0,m_{\text{cut}}] $ instead of the full range, where we take $m_{\text{cut}}= 1 $ in our simulations.  We check that the truncation procedure has not resulted in any change in the results by increasing the cutoff where the states are truncated, and checking the fidelity between the two results.  The truncation procedure is very effective in reducing the Hilbert space size and enables several orders of magnitude reduction.

\section{Results of numerical time evolution}

In this section we show our numerical results for our entangling procedure between the atomic ensembles. We first verify that it is possible to perform the STIRAP process adiabatically by analyzing the fidelity between the initial and target states without the presence of spontaneous emission and cavity loss.  We then examine the problem where both decoherence effects -- spontaneous emission and photon loss -- are included.

\subsection{Fidelity of adiabatic evolution}

\begin{figure}
\begin{center}
\includegraphics[width=\linewidth]{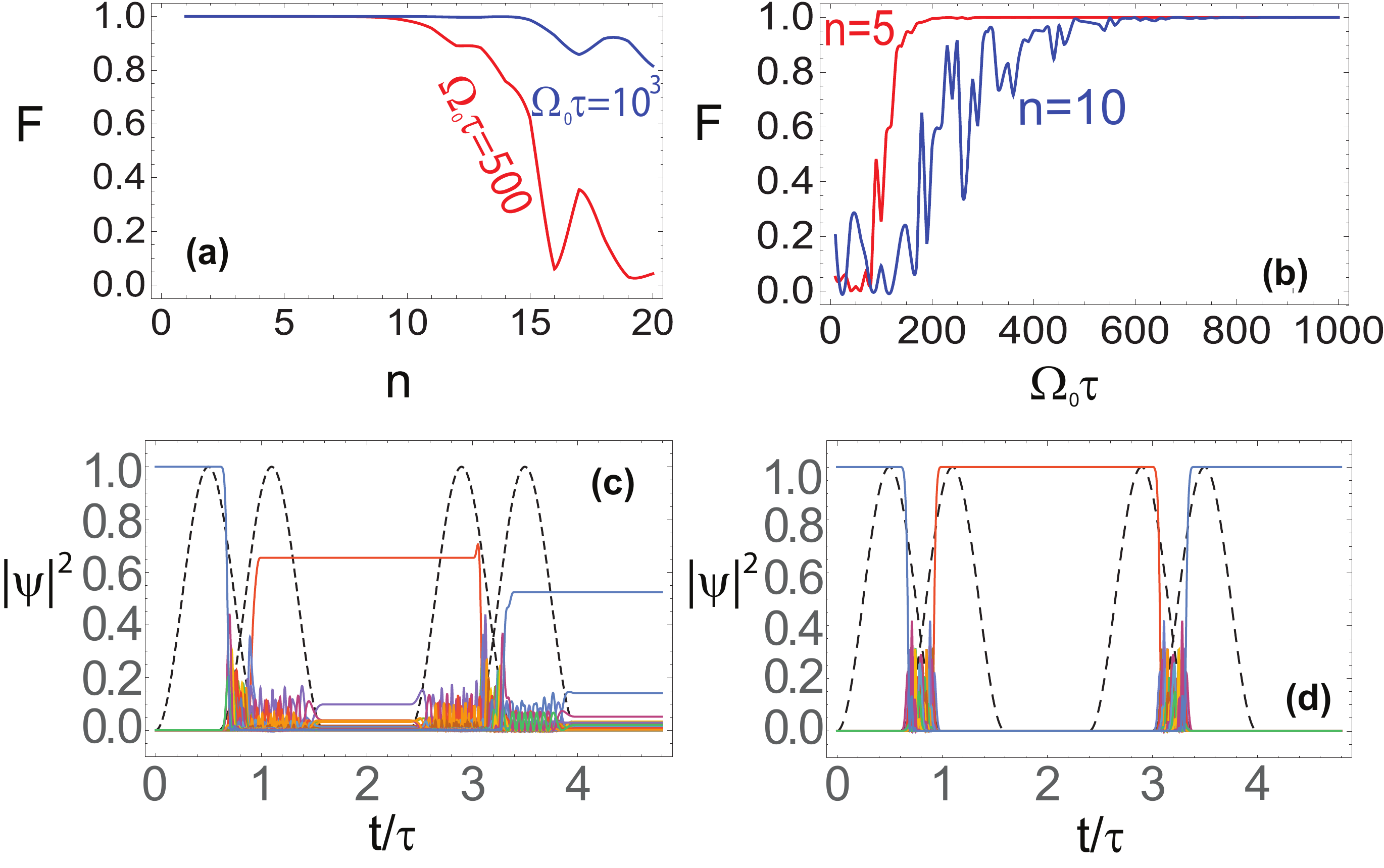} 
\caption{\label{fig5}
Performance of the STIRAP sequence with no spontaneous emission and no cavity loss.  (a)(b)
Fidelity between the initial state and final state $ F = |  \langle 0,0,0,0,0 | \psi(t=t_{\text{f}}) \rangle |^2 $.  Dependence on the (a) number of atoms $ n = n_1 = n_2 $ and (b) the amplitude of the STIRAP pulses $ \Omega_0 \tau $ are shown.   (c)(d) Distribution of Fock states $ | \psi_{\bm{n}} |^2  $ as given in (\ref{statedef}) for $ n_1 = n_2 = 10 $. Solid lines show the various Fock states and dotted lines show the position of the pulses (amplitude is arbitrary). (c) Non-adiabatic case $ \Omega_0 \tau = 200 $; (d) adiabatic case $ \Omega_0 \tau = 10^3 $.  Common parameters used in the plots are $ \delta t/\tau = 0.6, \Delta T/\tau = 3, G_1 \tau=G_2 \tau = \Omega_0 \tau, \Delta_e  = 0 $.  }
\end{center}
\end{figure}

As discussed in Sec. \ref{sec:stirap} and Fig. \ref{fig1b}, after the first STIRAP pair the minimum of the number of atoms between levels $ a_1 $ and $ b_2 $ is transferred to each of the levels $ b_1 $ and $ a_2 $. The second STIRAP pair then reverses the evolution such that the initial state is again recovered.  To verify that this process is being performed adiabatically, we calculate the fidelity between the numerically evolved state and the theoretically predicted state,  given by
\begin{align}
F =\Big( \text{Tr} \sqrt{\sqrt{\rho(t= t_{\text{f}} )} \rho(t=0) \sqrt{\rho(t= t_{\text{f}} )}} \Big)^2 
\end{align}
where $  t_{\text{f}} $ is the time after the STIRAP pulses are complete. Since the Hamiltonian is block diagonal in terms of the total particle numbers in levels $ a_i, b_i, e_i $, we evolve just a particular subsector with particle number $ n_1, n_2 $.  The initial state is then chosen to be the Fock state
\begin{align}
\rho(t=0) = | 0,0,0,0,0 \rangle_{n_1 n_2} \langle 0,0,0,0,0 |_{n_1 n_2}  .
\end{align}

Figure \ref{fig5} shows typical results of the numerical evolution.  As we see from the simulations, generally high fidelities close to 1 are possible in most of the parameter range with suitable parameters. 
Fig. \ref{fig5}(a) shows the scaling of the fidelity with respect to the particle number $ n = n_1 = n_2 $.  On first glance, the scaling with $ n $ appears to be rather poor, with the fidelity generally dropping exponentially as the boson number is increased.  It should however be pointed out that for a larger system it is easier for the system to lose fidelity due to the larger number of states that are available. This is a natural consequence of using a larger system and has been seen to occur in similar situations  \cite{pyrkov14,Ilo-Okeke2014}.  The larger number of states also allows for potentially more types of states to be entangled, so does not necessarily signal that the scheme is intractable.  As can be seen in Fig. \ref{fig5}(c), in the non-adiabatic case there are a population of states that do not get returned to the original states.  Such states can potentially still contribute to entanglement between the systems. 

The poor scaling can be countered by increasing the amplitude of the STIRAP pulses $ \Omega_0 $, or equivalently increase the pulse duration $ \tau $.  As seen in Fig. \ref{fig5}(b), this has the effect of exponentially improving the fidelity.  We find that to achieve a similar fidelity, the increase with 
 $ \Omega_0 \tau $ is roughly linear with $ n $, due to both effects being exponential.  While there are always experimental bounds to what laser amplitude, cavity coupling, and duration can be achieved, the overall scaling with $ n $ appears to be effectively linear.  For fidelities close to 1, the population curves typically appear as Fig. \ref{fig5}(d), with a single Fock state before and after each STIRAP pair.

\subsection{Including spontaneous emission and cavity loss}

\begin{figure}[t]
\begin{center}
\includegraphics[width=\linewidth]{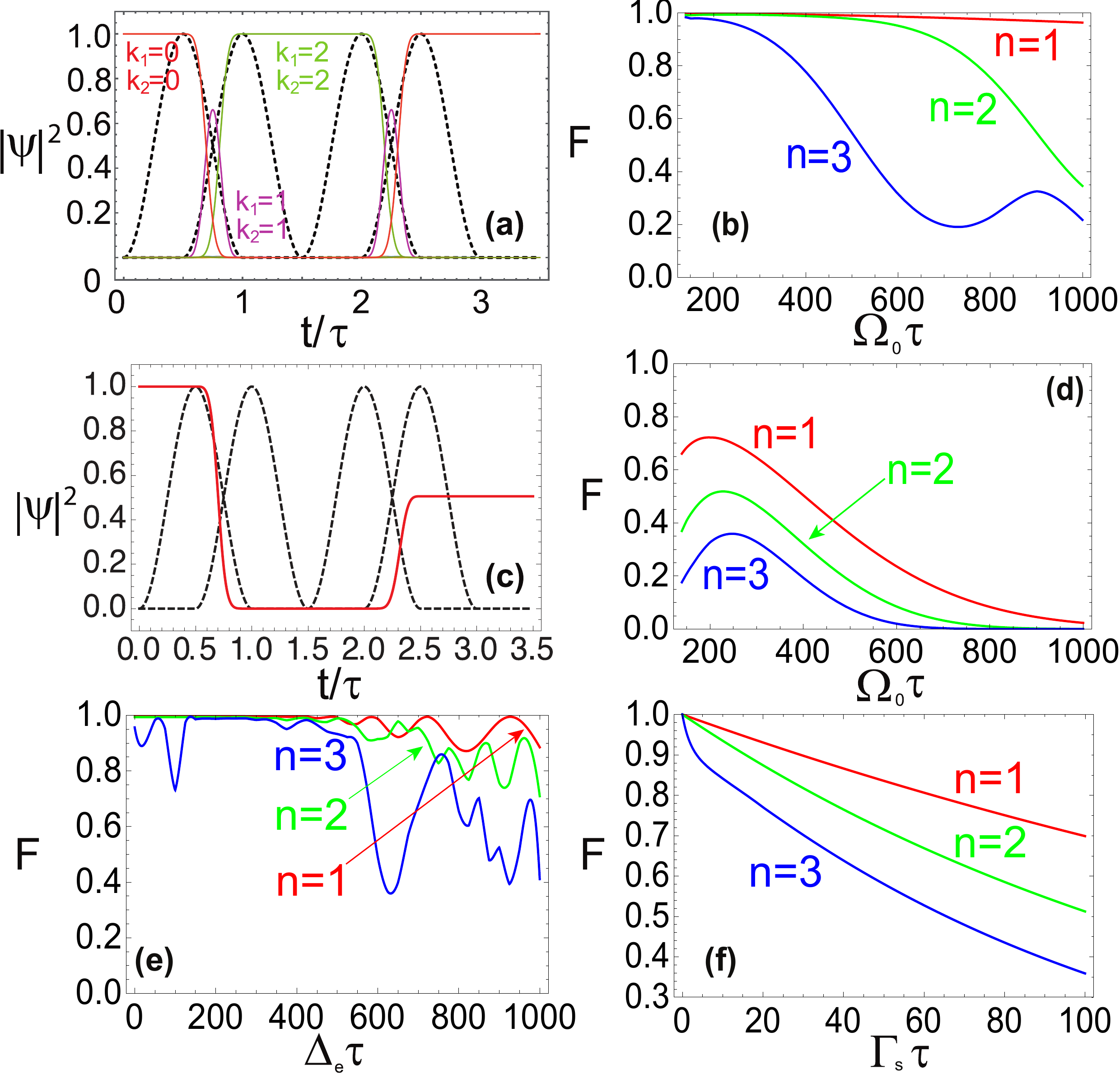} 
\caption{\label{fig6} 
Performance of the STIRAP sequence with various parameters. (a) Population of the adiabatic evolution for the case $ n_1=n_2 = 2 $. Labeled states have $ m_{1,2} = 0 $ and $ l = 0 $. All other unlabeled states have negligible population.  Parameters used are $ \Omega_0 \tau = 200 , \Gamma_s \tau=\Gamma_\gamma \tau= 0 , \Delta_e \tau = 0$. (b) Fidelity between the initial state and final state $F = | \langle 0,0,0,0,0 | \psi(t=t_{\text{f}})\rangle|^2$ with $ \Gamma_s \tau=\Gamma_\gamma \tau=1 , \Delta_e \tau = 0$.   (c) Population for the state $|0,0,0,0,0\rangle$ with $ n_1=n_2 = 2 $ with  $ \Omega_0 \tau = 200 , \Gamma_s \tau=\Gamma_\gamma \tau=100, \Delta_e \tau = 0 $.  (d)(e)(f) Fidelity between the initial state and final state $F = | \langle 0,0,0,0,0 | \psi(t=t_{\text{f}})\rangle|^2$.  Parameters used are  (d)  $  \Gamma_s \tau=\Gamma_\gamma \tau=100 , \Delta_e \tau = 0$; (e)  $ \Gamma_s \tau= \Gamma_\gamma \tau=1, \Omega_0 \tau=250$; and (f)  $ \Gamma_\gamma = \Gamma_s, \Omega_0 \tau=250, \Delta_e \tau = 0 $. The common parameters for all plots are $  G_1 \tau=G_2 \tau = 10^3, \delta t/\tau = 0.5, \Delta T/\tau =2.  $ }
\end{center}
\end{figure}

An optimum choice of the laser transition amplitude $ \Omega_0 $ exists such that both adiabaticity is maintained and the photonic population is suppressed.  According to Fig. \ref{fig5}(b) we see that in the interest of maintaining adiabaticity, generally a larger  $ \Omega_0 $ is desirable.  While this is effective in reducing the excited state population $ e_{1,2} $, this can result in a significant population of the cavity photon state.  This can be seen from the form of (\ref{qubitdarkd2}) for the qubit case, where the intermediate term $ |b_1 b_2 1 \rangle $ contains a cavity photon.  During the adiabatic transition cavity photon loss may occur which will contribute to decoherence.  To reduce the contribution of this term, one can choose $ \Omega_0 \ll G_{1,2} $ which reduces the weight of the intermediate terms.  Figure \ref{fig6}(a) shows an effective choice of parameters where both adiabaticity and suppression of the cavity photon population is maintained. Here we see that only the states with zero excited states and zero cavity photons have significant populations during the adiabatic evolution.  Such a parameter regime is most desirable for completing the adiabatic evolution with a high fidelity. Figure \ref{fig6}(b) shows the dependence of the final fidelity with $ \Omega_0 $ for various boson numbers with spontaneous emission parameters chosen as $ G/\Gamma_s = G/\Gamma_\gamma \sim 10^3 $.  For small values of  $ \Omega_0 $ we see good performance, with near unit probabilities of returning to the ground state.  

We find that the parameters necessary to obtain high probability population transfer back to the ground state is in the vicinity of  $ G/\Gamma_s = G/\Gamma_\gamma \sim 10^3 $.  Unfortunately this exceeds typical cavity parameters, which are in the vicinity of $ G/\Gamma_s = G/\Gamma_\gamma \sim 10 $ (see for e.g. Ref. \cite{colombe07}).  Figure \ref{fig6}(c)(d) shows the time dependence of the population of the initial state and final fidelities with parameters $ G/\Gamma_s = G/\Gamma_\gamma = 10 $.  As expected the performance is degraded considerably, as would be expected by including decoherent processes.  We observe that an optimum value of $ \Omega_0 \tau \approx 250$ is present for various boson numbers $ n $, in accordance with an optimal value that both suppresses the photonic and excited state populations.  

In  Fig. \ref{fig6}(e) we show the dependence of the fidelity with the detuning $ \Delta_e $. We find that introducing detuning does not effectively improve the fidelity, with poor performance being attained for large values.  We therefore focus on  the optimal parameters
$ \Omega_0 \tau \approx 200 $ and zero detuning unless otherwise stated.  In Fig. \ref{fig6}(f), we show the dependence of the fidelity on the spontaneous emission rate.  As expected, the fidelity decreases with $ \Gamma_s $.  We observe that the rate of decrease is larger for larger boson numbers.  This is also expected due to the bosonic enhancement effects of decoherence with large boson number.  The scaling with $ n $ appears to be rather poor, with a decoherence rate scaling approximately linearly.  We thus observe that despite using the STIRAP scheme it is still difficult to suppress it effectively for larger boson numbers.  This appears to be a poor result given that in realistic BECs one will have typically $ n > 100 $.     We shall however see in the following section that, despite the poor fidelity scaling, surprisingly robust entanglement still can be present in the system.

\begin{figure*}
\includegraphics[width=\linewidth]{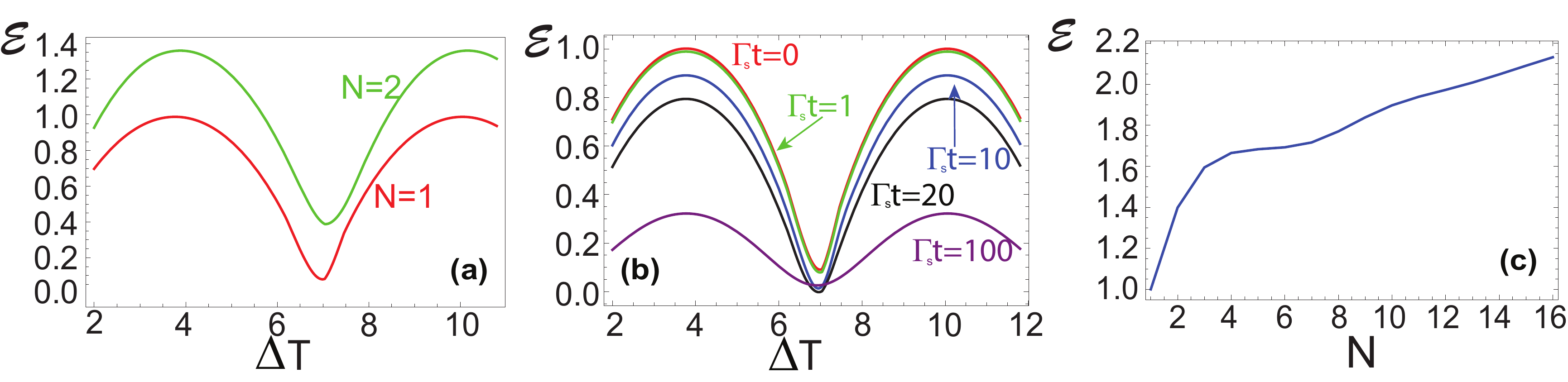} 
\caption{
\label{fig7}
The logarithmic negativity (\ref{negativitydef}) for the STIRAP sequence including spontaneous emission and cavity photon loss  with (a) $ \Gamma_s \tau= \Gamma_\gamma \tau= 1, \Delta_e \tau=100 $; (b) $ N_1 = N_2 = 1, \Gamma_\gamma=\Gamma_s, \Delta_e   \tau=100 $; (c) $ N = N_1 = N_2, \Gamma_s \tau= \Gamma_\gamma \tau= 0, \Delta_e   \tau= 0 $ and  $ \Delta T$ is optimized such that the maximal entanglement is given.   Common parameters used are  $\Omega_0 \tau = 200, \delta t/\tau = 0.5,  G_1 \tau=G_2 \tau =1000, \omega \tau=1, m_{\text{cut}}= 1 $.   }
\end{figure*}

\subsection{Entanglement}
\label{sec:level2}

We now calculate the entanglement generated by the scheme. In order to generate entanglement, a time dependent phase $ e^{-i \omega t} $ is required in (\ref{pulsedef}) to add an entangling phase to the STIRAP evolution. We start from the 
state (\ref{initcondition}) and evolve the full density matrix including spontaneous emission and cavity loss. The entanglement is calculated using the logarithmic negativity is defined as 
\begin{equation}
{\cal E}=\log_{2} || \rho^{T_{1}}||=\log_{2}\sum_{i}|\lambda_{i}|
\label{negativitydef}
\end{equation}
where $\rho^{T_{1}}$ is the partial transposed density matrix with respect to ensemble 1, the $ || \cdot || $ takes the trace norm of a matrix, and $\lambda_{i}$ are the eigenvalues of $\rho^{T_{1}}$. The logarithmic negativity is a sufficient condition for entanglement, meaning that a non-zero value guarantees entanglement is present, whereas a zero value is inconclusive \cite{vidal2002computable}.    Previous studies with similar types of entanglement have shown that the logarithmic negativity gives qualitatively similar results as the von Neumann entropy, and has the same maximum value of 
$ {\cal E}_{\text{max}} = \log_2 (N+1) $ \cite{byrnes13}.  

Due to the large number of levels involved (see Fig. \ref{fig1b}) we were only able to simulate the entanglement directly for relatively small boson numbers $ N_{1,2} \le 2 $.  Fig. \ref{fig7}(a) shows our results for the optimal parameters as discussed in the previous section.  We see that generally the same behavior as Fig. \ref{fig2} is obtained, where the negativity has a periodic structure with a periodicity that is controlled by $ \omega $. In Fig. \ref{fig7}(b) we calculate results for cavity parameters corresponding to experimentally achieved values in the range  $ G/\Gamma_s = G/\Gamma_\gamma \sim 10 $.  We see that this surprisingly has a rather good scaling with larger values of decoherence.  Even for values that are $ 100 \times $ the limit where one would obtain good results based on the results of fidelity, one obtains significant amounts of entanglement.  We attribute this to the fact that states other than those intended by the scheme (i.e. $ a_1 $ and $ c_1 $; $ b_2 $ and $ c_2 $) can contribute to the entanglement.  The decoherence terms can result in an inadvertent population of these other states, which are counted in the negativity calculation.  

In order to explore larger values of $ N_{1,2} $ we use an approximate scheme to verify that the correct behavior to the entanglement is indeed generated by the scheme.  Evolving the pure state (\ref{schrodinger}) requires far less resources than evolving the density matrix directly and larger values can be calculated.  Our procedure is to start from the state (\ref{initcondition}) and calculate the Berry phase numerically for each of the Fock states in the expansion (\ref{spincoherentstate2}) and (\ref{spincoherentstate4}).  This (pure) state is then substituted into (\ref{negativitydef}) to obtain the negativity. Fig. \ref{fig7}(c)  shows our results.  It shows the scaling of the entanglement with respect to the boson number $N = N_1 =N_2$. The effective linear scaling when $N$ is large shows that the entanglement procedure also works well for large boson systems. The difference with poor scaling in Fig. \ref{fig6}(d) can be explained by again the contribution of states that don't return to the original state, but still contribute to the entnaglement.  We see that the same general behavior is obtained as in Fig. \ref{fig2}(a) with a logarithmic increase in negativity. 

While an explicit calculation of the entanglement for large $ N_{1,2} $ is difficult due to the numerical overhead, the results of Fig. \ref{fig7}(b) for large values of decoherence are encouraging due to the general expectation that decoherence effects are enhanced for larger boson numbers. Generally due to superradiance, spontaneous emission is enhanced by a factor of $ N $ due to bosonic enhancement.  The cavity photon loss on the other hand is not enhanced because we always work in a regime where the cavity photon population is small.  In this regard, it is more important to overcome the spontaneous emission, which the STIRAP is effective at doing.  As long as the cavity photon population is suppressed to levels as that shown in Fig. \ref{fig7}(a)(b), we expect that the scheme can produce entanglement even in the case of a large number of atoms.  For the best results the scheme apparently requires rather good cavities with parameters in the range  $ G/\Gamma_s = G/\Gamma_\gamma \sim 10^3 $.

\section{Summary and conclusions}
\label{sec:conc}

We have proposed a method for entangling two ensembles using an adiabatic evolution involving a common cavity mode to mediate the interaction. The scheme possesses a dark state for all particle numbers in the cavities, including the case $ N_1 = N_2 = 1 $, which reduces to a qubit case. While we have been primarily concerned with creating entanglement between ensmebles, this can be equally be applied for standard qubits linked by a common cavity mode. 
 The presence of the dark state allows for an adiabatic evolution to produce a geometric phase gate for superposition states between ground states of the atoms.  The geometric phase produced by the adiabatic evolution has an unusual form, depending on the minimum of the number of atoms on one of the logical states on each of the ensembles.  One of the main benefits of using the dark states is that it helps to overcomes spontaneous emission, one of the main decoherence channels for schemes using excited states in ensembles.  We find that in our numerical simulations on small systems that it is possible to generate significant amounts of entanglement, even in the presence of spontaneous emission and cavity photon loss.  The best results are obtained for cavities with very strong coupling, where  $ G/\Gamma_s = G/\Gamma_\gamma \sim 10^3 $, but entanglement is still produced in cavities with parameters in the currently realizable range $ G/\Gamma_s = G/\Gamma_\gamma \sim 10 $.   The key to this is to use an optimized laser amplitude which works in an adiabatic regime, but has $ \Omega_0 \ll G_{1,2} $, which suppresses the cavity photon population.  One of the difficulties we encountered was the numerical complexity of simulating the system for large ensemble populations. While this prevented us from directly simulating the entanglement, the results  of Fig. \ref{fig7}(b) are encouraging, as significant amounts of entanglement are present even for imperfect adiabatic transitions.  Alternative numerical methods based on stochastic evolution \cite{gardiner2004quantum} may be a way to improve on the numerical results given here.  

The form of the effective interaction Hamiltonian $H_{\text{eff}} =\min(k_{1},k_{2})$ is interesting not only from an entanglement point of view, but also for computational purposes.  The minimum operation is a key operation in constraint programming 
used in sophisticated algorithsms such as next algorithm search, next greater element, or cycle algorithm \cite{constraintA,constraintB}.   These are equivalent to the Hamiltonian path problem \cite{NP_Problem}, which is a type of NP-complete problem. The minimum operation is one of the necessary algorithms to solve such problems.  One possible way that our gate could be used in this context is to use the logical states of the ensemble as a quantum register, after which our proposed scheme could be applied to calculate the minimum with quantum parallelism.  This may be incorporated as a logical primitive for optimization problems, and may be applicable to problems such as quantum machine learning \cite{biamonte2016quantum}.

\acknowledgments

We would like to thank Marek Narozniak, Joan Vazquez, Daniel Rosseau, and Sandrine Idlas for useful discussions. S.O. thanks the National Institute of Informatics (NII) for its International Internship Program. This work is supported by the Shanghai Research Challenge Fund; New York University Global Seed Grants for Collaborative Research; National Natural Science Foundation of China (61571301); the Thousand Talents Program for Distinguished Young Scholars (D1210036A); and the NSFC Research Fund for International Young Scientists (11650110425); NYU-ECNU Institute of Physics at NYU Shanghai; the Science and Technology Commission of Shanghai Municipality (17ZR1443600); and the China Science and Technology Exchange Center (NGA-16-001).

\bibliographystyle{apsrev}
\bibliography{references}

\end{document}